\documentclass[aps,showpacs, prl,twocolumn,superscriptaddress]{revtex4}%
\usepackage{epsfig,dsfont,amssymb,amsmath,amsthm,amsfonts,amsbsy,mathrsfs}
\usepackage{graphicx}
\usepackage{amsmath}
\usepackage{amssymb}
\begin{document}
\title{Nano-Stitching of Graphene Bilayers: A First-Principles Study}
\author{Chaoyu He}
\affiliation{Hunan Key Laboratory for Micro-Nano Energy Materials
and Devices, Xiangtan University, Hunan 411105, China}
\affiliation{School of Physics and Optoelectronics,
XiangtanUniversity, Xiangtan 411105, China}
\author{Jin Li}
\affiliation{Hunan Key Laboratory for Micro-Nano Energy Materials
and Devices, Xiangtan University, Hunan 411105, China}
\affiliation{School of Physics and Optoelectronics,
XiangtanUniversity, Xiangtan 411105, China}
\author{Jianxin Zhong}
\email{jxzhong@xtu.edu.cn}\affiliation{Hunan Key Laboratory for
Micro-Nano Energy Materials and Devices, Xiangtan University, Hunan
411105, China} \affiliation{School of Physics and Optoelectronics,
XiangtanUniversity, Xiangtan 411105, China}
\date{\today}
\pacs{61.46.-w, 61.72.J-, 66.23.St, 62.23.Kn}

\begin{abstract}
A nano-stitching method is proposed and investigated to modify graphene bilayers.
Based this method, four types of low energy carbon allotropes, "wormhole graphene" allotropoes,
are obtained and their structures, stabilities and electronic properties are investigated using
first principles methods. We find that all of these wormhole graphene allotropoes are more
favorable than graphdiyne and dynamically stable. Similar to carbon nanotubes and fullerences,
these graphene allotropes are expected to act as two-dimensional periodic nano-capsules for
encapsulating magnetic atoms or functional clusters for a variety of applications.\\
\end{abstract}
\maketitle \indent Graphene \cite{1,2} is a single layer of carbon
atoms in two-dimensional hexagonal lattice. It has attracted
tremendous attentions owing to its peculiar electronic structures
\cite{2,3,4,5} since it was synthesized in 2004 \cite{1, 2}. In the
past decade, graphene layer has always been considered as
nano-fabrics and geometrically cut into multifarious patterns
\cite{6,7,8} for the purposes of designing special functional
segments using in integrated circuit. Electronic and magnetic
properties of graphene can be efficiently modulated by such a
geometrical cutting method \cite{6,7,8}. For examples, graphene
nanoribbons with zigzag and armchair edge shapes show different
electronic and magnetic properties \cite{6}. Graphene layer has also
been considered as nano-paper and chemically painted in different
patterns with different functional atoms and/or molecules
\cite{9,10,11}. For example, a new type of 2D material graphane was
theoretically proposed \cite{12,13,14,15,16,17,18} and
experimentally synthesized through exposing graphene to cold
hydrogen plasma \cite{10, 11}. These two main types of modification methods provide
us efficient modulating effects on the physical properties of graphene.\\
\indent In fact, graphene layer has also been considered as the building blocks to form carbon nanotubes, fullerences
and graphite. In this work, we theoretically propose a nano-stitching method to modify graphene bilayers for designing
a new type of carbon-based material, which we named it as "wormhole graphene". Four stitching manners of graphene
bilayers are considered and investigated, which have perfect interlayer links, C1 vacancy based interlayer links,
C4 vacancy based interlayer links and C6 vacancy based interlayer links, respectively. Based on the 4x4 supercells
of AA stacked graphene bilayer, four corresponding wormhole graphene allotropes are obtained and their structures,
stabilities and electronic properties are investigated. We find that all of these wormhole graphene allotropes are
more favorable than graphdiyne and dynamically stable. Similar to carbon nanotubes and fullerences, these graphene
allotropes are expected to act as 2D periodic nano-capsules for encapsulating magnetic atoms or clusters for different
applications.\\
\section{Computational Details}
\indent The nano-stitching method is schematically shown in Fig.1. The wormhole-like graphenes can be obtained through three steps:
I). Stack two graphene layers in an AA-stacking manner with or without intercalated atoms/functional-clusters. II) Periodically etch
the graphene bilayer into a bilayer nano-mesh with holes in different sizes, creating active dangling atoms at the edge of the holes.
III). Compress the graphene bilayer nano-mesh to induce a reconstruction of the dangling atoms in different layers, "stitching" the two
graphene layers together to form the wormhole graphene. Such a nano-stitching method can also be applied to h-BN bilayer and graphene/h-BN
bilayer with or without intercalated functional atoms or clusters. In this work, we use the graphene bilayer without intercalated atoms/clusters
to demonstrate the nano-stitching technology. Four types of stitching manners between graphene layers are considered and investigated, which have
perfect interlayer links (P-IL), C1 vacancy based interlayer links (C1-IL), C4 vacancy based interlayer links (C4-IL) and C6 vacancy based interlayer
links (C6-IL), as shown in Fig.2. Based on the 4x4 supercells of AA stacked graphene bilayer, four corresponding wormhole graphene allotropes are obtained
and their structures, stabilities and electronic properties are investigated.\\
\begin{figure}
\includegraphics[width=3.5in]{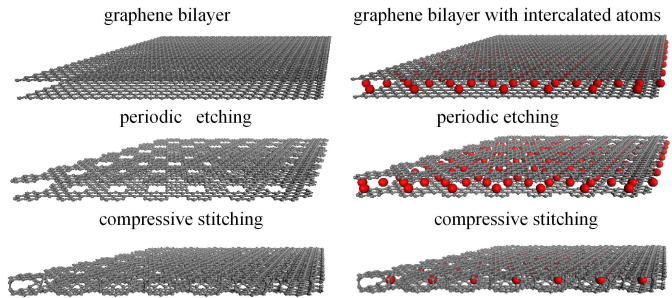}
\caption{The schematic of the nano-stitching method as implemented
in graphene bilayer (left) and graphene bilayer with intercalated
functional atoms/clusters (right).}\label{fig1}
\end{figure}
\indent Our calculations were carried out using density functional
theory with local density approximation \cite{19, 20} (LDA) as
implemented in Vienna ab initio simulation package (VASP) \cite{21,
22}. The interactions between the nucleus and valence electrons of
carbon atoms were described by the projector augmented wave (PAW)
method \cite{23, 24}. A plane-wave basis with a cutoff energy of 500
eV was used to expand the wave functions of all systems. The
Brillouin Zone (BZ) sample meshes for all systems were set to be
dense enough in our calculations (5x5x1). Before property
calculations, we optimized lattice constants and atomic positions
for all the systems through the conjugate-gradient algorithm until
the residual force on every atom is less than 0.01 eV/A. The
dynamical stabilities of these wormhole graphenes were evaluated by
simulating their vibrational properties through the phonon package \cite{25} with the forces calculated from VASP.\\
\begin{figure}
\includegraphics[width=3.5in]{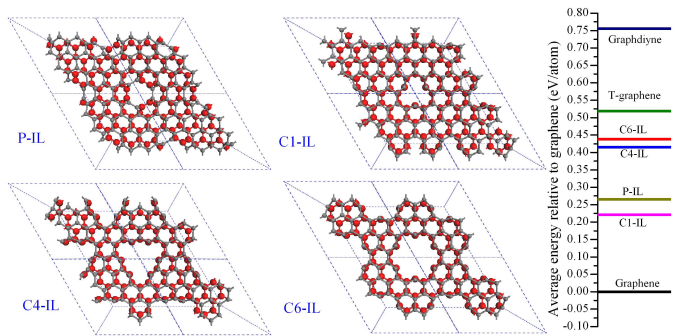}
\caption{The perspective views of the wormhole graphene with P-IL,
C1-IL, C4-IL and C6-IL interlinks and their corresponding average
energies respective to that of graphene.}\label{fig2}
\end{figure}
\section{Results and Discussions}
\indent We first discuss the structural properties of these wormhole
graphene allotropes. All the interlink junctions between graphene
bilayers are constructed to be all-sp$^2$ configurations. After full
optimization, the configurations maintain the all-sp$^2$. One can
reproduce the crystal structures of the wormwhole graphene
allotropes based on their crystalline information as discussed below
and see the details about the interlink junctions in any 3D view
software. Crystal structure of P-IL belongs to the space group 162
(P-3M1) and possesses lattice constants of a=b=9.87 {\AA} and c=18
{\AA}. It contains 64 carbon atoms in its crystal cell but only six
of them are nonequivalent. The six nonequivalent atomic positions
are (0.848,0.769,0.536), (0.819,0.652,0.589), (0.661,0.576,0.611),
(0.915,0.585,0.602), (0.578,0.411,0.617) and (0.667,0.333,0.618),
respectively. In P-IL, the interlayer chemical bonds are formed by
12 junction atoms from different layers. A -6-8-6-8- carbon rings
sequence can be noticed in the junction area. C1-IL possesses the
P-6M2 (187) symmetry and its lattice constants are a=b=9.732 {\AA}
and c=18 {\AA}. It contains 62 carbon atoms per crystal cell but
most of them are equivalent. There are nine nonequivalent atomic
positions in C1-IL. They are (0.897, 0.794, 0.539), (0.747, 0.746,
0.577), (0.667, 0.834, 0.581), (0.501, 0.706, 0.596), (0.416, 0.583,
0.599), (0.499, 0.500, 0.603), (0.667, 0.584, 0.599), (0.750, 0.500,
0.603) and (0.667, 0.333, 0.606). The interlayer bonds in C1-IL are
formed by 6 junction atoms from different layers, forming a
-10-10-10- carbon rings sequence. C4-IL possesses the P-6M2 (187)
symmetry and its lattice constants are a=b=9.713 {\AA} and c=18
{\AA}. The crystal cell of C4-IL contains 56 carbon atoms
distributed in eight nonequivalent atomic positions. The eight
nonequivalent positions are (0.707, 0.729, 0.541), (0.654, 0.827,
0.573), (0.502, 0.751, 0.611), (0.517, 0.835, 0.611), (0.501, 0.001,
0.597), (0.418, 0.082, 0.578), (0.751, 0.249, 0.578) and (0.667,
0.333, 0.587). The 12 junction atoms around the C4 vacancy bond to
each other to form the junction with a -6-10-6-10-6-10- carbon rings
sequence. For C6-IL, there are 52 carbon atoms in its hexagonal cell
with lattice constants of C6-IL are a=b=9.782 {\AA} and c=18 {\AA}.
It belongs to the P6/MMM (191) space group and possesses only four
nonequivalent carbon atoms locating at (0.817, 0.635, 0.538),
(0.665, 0.584, 0.575), (0.417, 0.583, 0.587) and (0.667, 0.333,
0.586). In C6-IL, the interlayer junction with a -8-8-8- carbon rings sequence is formed by 12 carbon atoms from different layers.\\
\indent The relative thermodynamic stabilities of these wormhole graphene allotropes were evaluated through comparing their average
energies to those of graphene, T-graphene and graphdiyne. As shown in Fig.2, we can see that C1-IL is the most stable one among
these four types of stitching manners. Although all of total energies of these wormhole graphene allotropes are higher than that
of graphene, all of them are more favorable than the theoretically proposed T-graphene and the experimentally synthesized graphdiyne.
Dynamical stabilities of these wormhole grpahenes were also evaluated through calculating their vibrational properties. As shown in
Fig.3, we plot the calculated phonon density of states of the wormhole graphene with P-IL, C1-IL, C4-IL and C6-IL interlinks, respectively.
We can see that there are not any vibrational modes in the imaginary frequency area in the whole Brillouin Zone, which confirms that all these
wormhole graphene allotropes are dynamically stable. In view of the fact that graphdiyne has been experimentally synthesized, we expect these
energetically more viable wormhole graphene allotropes can also be synthesized in the future through the nano-stitching method as discussed above.\\
\begin{figure}
\includegraphics[width=3.50in]{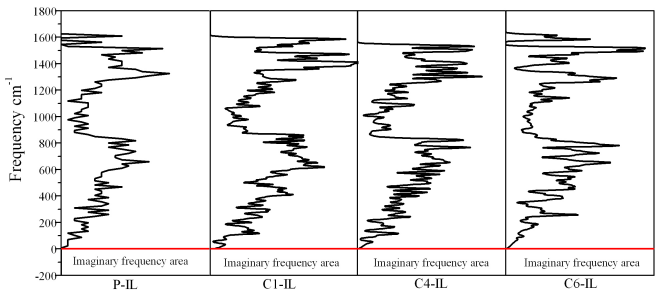}
\caption{Calculated phonon density of states of wormhole graphene
with P-IL, C1-IL, C4-IL and C6-IL interlinks,
respectively.}\label{fig3}
\end{figure}
\indent After the discussion about the structures and stabilities of the wormhole graphene, we turn our attention to the electronic properties
of these new carbon allotropes. The calculated electronic band structures of the four wormhole graphene allotropes are shown in Fig.4. The band
structure of a 4x4 AA-stacked bilayer graphene is also inserted for comparison. We can see that the P-IL possesses dirac-like property similar
to that of the AA-stacked bilayer graphene around Fermi-level. But the band branches far from the Fermi-level are very different in P-IL and
AA-stacked bilayer graphene, which indicates that interlayer bonds have strong modulation effects on the electronic properties. Furthermore,
the dirac-cone in P-IL is located a little below the Fermi-level, which indicates that the interlayer bonding roles as n-type doping. The band
structure of C1-IL shows obvious difference in comparison to that of the perfect AA-stacked bilayer graphene. The appearance of C1 vacancy and
the interlayer bonding between vacancy atoms induce obvious p-type doping effect in the band structure, which moves the dirac-cone a little above
the Fermi-level and makes C1-IL show metallic property. From the band structures we can see that such a p-type doping effect is more obvious in C4-IL
and C6-IL. Both C4-IL and C6-IL can be confirmed as metals according to their band structures.\\
\indent In our present work, only the situation of a 4x4 AA-stacked graphene bilayer without any intercalated atom/functional-clusters is considered
to demonstrate our nano-stitching method. In fact, such a nano-stitching method can also be applied to h-BN bilayer and graphene/h-BN bilayer with or
 without intercalated functional atoms or clusters. We believe that the nano-stitching method can be experimentally implemented in the future and it
 can provide us abundant 2D wormhole-like materials with different properties through changing the type of the bilayer, the species of the intercalated
 atoms/clusters, and the periodicity and size of the hole. Further theoretical and experimental efforts in this direction are expected to help to design
 and synthesize new 2D periodic nano-capsules with various encapsulated magnetic atoms or clusters for special applications.\\
\begin{figure}
\includegraphics[width=3.50in]{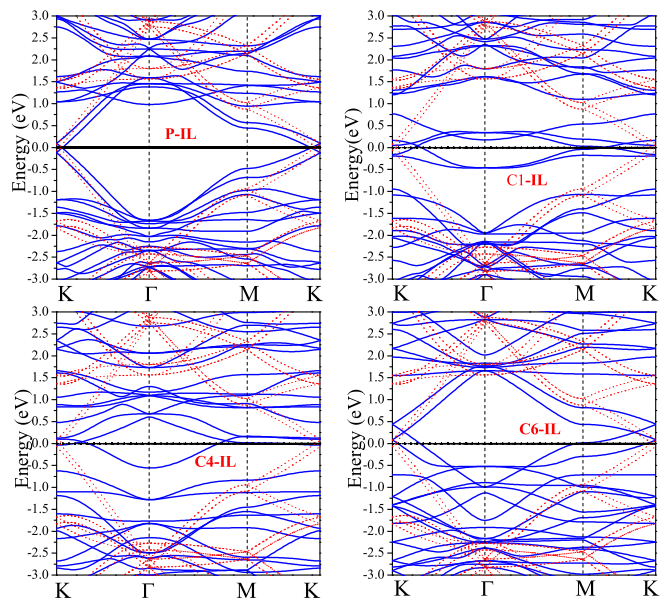}\\
\caption{Calculated electronic band structure of the wormhole
graphene with P-IL, C1-IL, C4-IL and C6-IL interlinks,
respectively.}\label{fig4}
\end{figure}
\section{Conclusion}
\indent We have proposed a nano-stitching method to modulate structures and properties of graphene and other 2D materials.
A 4x4 AA-stacked graphene bilayer has been considered as an example to demonstrate the method with four stitched wormhole
graphene allotropes of distinct interlayer connections. The structures, stabilities and electronic properties of the four
wormwhole graphene allotropes were investigated by first-principles calculations. We found that all of these wormhole graphene
allotropes are more favorable than graphdiyne and dynamically stable, indicating that the nano-stitching method applicable.
We expect that further theoretical and experimental efforts in this direction can help to design and synthesize new functional
2D periodic nano-capsules with various encapsulated magnetic atoms or clusters for special applications.\\
\section{Acknowledgements}
This work is supported by the National Natural Science Foundation of
China (Grant Nos. A040204 and 11204261), the National Basic Research
Program of China (2012CB921303), the Hunan Provincial Innovation
Foundation for Postgraduate (Grant No. CX2013A010), and the Young
Scientists Fund of the National Natural Science Foundation of China
(Grant No. 11204260).

\end{document}